\documentclass[aip,apl,groupaddress,showkeys,twocolumn]{revtex4}
\usepackage{graphicx}
\begin{document}
\newcounter{roman3}
\newcounter{roman5}
\setcounter{roman3}{3}
\setcounter{roman5}{5}

\title{Patterned backgating using single-sided mask aligners: application to  density-matched electron-hole bilayers}

\author{A.F. Croxall}\author{K. Das Gupta}\email{kd241@cam.ac.uk}\author{C.A.Nicoll} \author{M. Thangaraj} \author{I. Farrer} \author{D. A. Ritchie} \author{M. Pepper}
\affiliation{Cavendish Laboratory, University of Cambridge, J.J. Thomson Avenue, Cambridge CB3 0HE, UK.}

\begin{abstract}
We report our work on fabricating lithographically aligned patterned
backgates on thin (50-60$\mu$m) \Roman{roman3}-\Roman{roman5} semiconductor samples using {\it single sided mask
aligners only}. Along with this we also present a way to photograph
both sides of a thin patterned chip using  inexpensive infra-red
light emitting diodes (LED) and an inexpensive (consumer) digital camera.
A robust method of contacting both sides of a
sample using an ultrasonic bonder is described. In addition we present
a mathematical model to analyse the variation of the electrochemical
potential through the doped layers and heterojunctions that are
normally present in most GaAs based devices. We utilise the
technique and the estimates from our model to fabricate an
electron-hole bilayer device in which each layer is separately
contacted and has tunable densities. The electron and hole layers are
separated by barriers either 25 or 15nm wide. In both cases, the densities can be
matched by using appropriate bias voltages.
\end{abstract}

 \maketitle
\section{Introduction}
Patterned backgating is a critical step in the processing of a large
class of devices that use double quantum well
structures,\cite{dblqwlist} or where manipulation of the shape and
position of the wavefunction at a heterointerface is envisaged
\cite{yablonovitch}. It is also of importance in devices where the
top surface needs to remain non-metallised for STM/AFM or optical
studies \cite{rolf}. Investigation of surface states of
semiconductors also requires a metal-free top surface, with a backgate
to change densities when required\cite{kawaharazuka}. Backgates can
be either lithographically aligned to topside features using
mask-aligners with double (bottom) side alignment capability or
patterned in-situ during wafer-growth by using Focused Ion-Beam
(FIB) techniques. \cite{fib} Both these solutions require complex
and very expensive additional equipment. In this paper we present a
technique of reproducible thinning of the sample (wet etching to
~50$\mu$m) and lithographically aligning a backgate with any existing
topside features using a single sided mask-aligner at all stages.
Conventional ultrasonic ball bonders can be used to contact gates on either
side of the sample. Additionally, the alignment can be verified and
photographed using inexpensive infra-red (IR) (wavelength 880/950nm) LEDs and a
standard inspection microscope fitted with a digital camera. We have
used this technique to study a 2-dimensional hole gas at an inverted
GaAs/AlGaAs interface and independently contacted electron-hole
bilayers in a double quantum well structure, however our technique
can be easily adapted for use with almost any other material. Other
techniques of fabricating lithographically patterned backgates ({\it
e.g.} EBASE: Epoxy Bond and Stop Etch)\cite{ebase} exist - however
the utility of the technique presented here lies in the fact that a
commonly available single-sided mask aligner can be used at all
stages.

The spatial variation of the bandstructure and the electrochemical
potential near an interface must be analysed before attempting to
change charge densities by backgating. We have used a self-consistent
Poisson-Schr\"{o}dinger-current equation solver
\cite{nextnano} to analyse the variation of the Fermi-level under
biasing and doping to estimate acceptable dopant densities without
causing pinning of the Fermi-level and undesirable hopping conduction.
Wafers were grown using these estimates as a guide and the patterned
backgating method described in the next section was applied to
fabricate  independently contacted electron-hole bilayer devices
with narrow barriers of the order of excitonic Bohr radius of GaAs
($\approx$12 nm).

\section{Method}
\begin{figure}[h]
\includegraphics[width=9cm,clip]{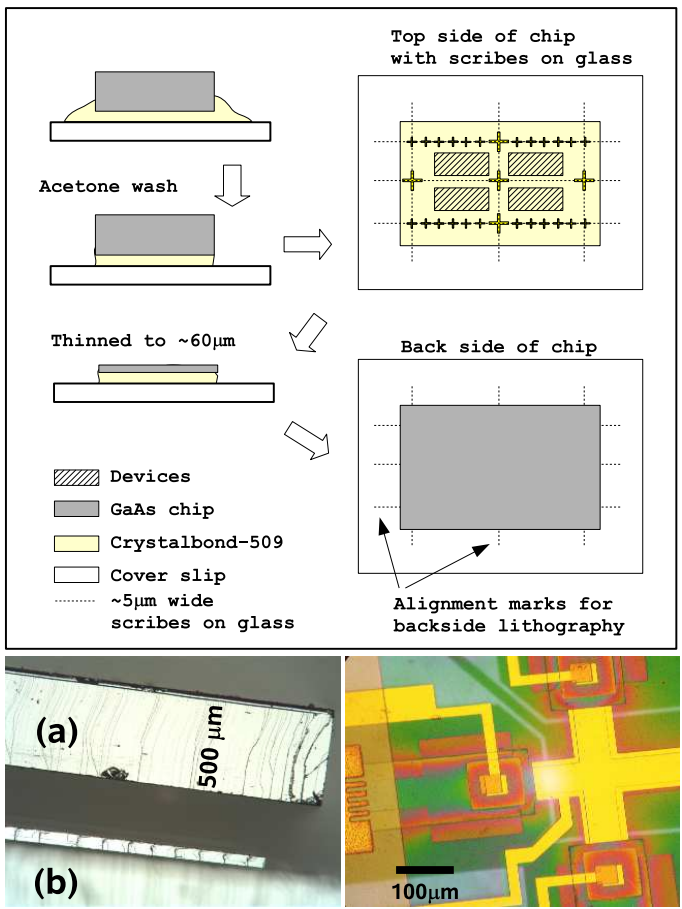}
\caption{\label{process-steps-fig} (Color online) Top: Figure
illustrates the processing steps for doing backside lithography.
Bottom left: typical cross-section of a device before (a)  and after (b)
thinning to $<$60 $\mu$m. Bottom right: Infra-red+visible composite photograph of a
processed device, taken using the method described in the section,
showing the alignment accuracy (better than $\sim 5\mu$m) that can be
obtained.} 
\end{figure}

\begin{figure}[h]
\includegraphics[width=9.0cm,clip]{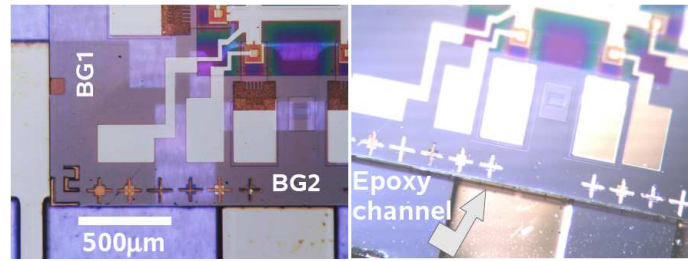}
\caption{\label{samplemountingfig} (Color online) Mounting of sample
for bonding to backgates. Left: Alignment of the backgate with
etched channels. Right: The etched channels for contacting the
backgates are about $\sim 20\mu$m deep and would be filled with a
small amount of Silver-epoxy.} 
\end{figure}

In our design each chip contains a few crosshair shaped alignment
marks on the topside at the centre and near the edges (see fig
\ref{process-steps-fig}). The wafers are originally 500$\mu$m thick.
On completion of topside processing the
chips are mounted topside down on a thin glass slide with a
transparent wax (Crystalbond \texttrademark 509 from SPI
technologies) The wax melts at around 130$^o$C, is resistant to the
acid peroxide etch solution used, but would dissolve in acetone if
soaked for a few hours. Using a diamond tipped mechanical scriber,
the (crosshair shaped) alignment mark is then copied and extended onto
the other side of the glass slide. It is for this step that the
transparency of the wax is crucial. The fine groove ($\sim$
5$\mu$m width) made on the glass are visible from both sides in
regions not covered by the chip. Thinning of the chip is then carried out using
an acid-peroxide (1:1:8 H$_2$SO$_4$: H$_2$O : H$_2$O$_2$) solution
with 2000:1 Triton-X surfactant (5ml per 50ml of etch solution). The
solution is continuously agitated with a magnetic stirrer during the
process. After an estimated time the chips are taken out and the
thicknesses of the chips are  measured with a step profilometer. The process is
continued untill the chips are 50-60$\mu$m thick. 
Lithography for the backgate layer is then done using the extended
grooves on the glass as  alignment marks, and a single
sided mask aligner. The alignment can be verified before
metallisation, using an inspection microscope fitted with a digital
camera. To view both sides of the chip simultaneously we need
radiation of a frequency to which a GaAs wafer is transparent.
This is very easily provided by an infra-red LED (880/950nm) placed
below the microscope stage. Since the CCD sensors used by all
digital cameras have a finite sensitivity to IR, the camera sees
both surfaces of the chip simultaneously (see fig\ref{process-steps-fig},\ref{samplemountingfig}) The strength of
the IR illumination from the bottom and visible light from the top
are adjusted to get a picture with good clarity of both sides.
After the backgate metallisation is done the individual devices can
be diced while the chips are still fixed on the glass. Finally a
long soak in acetone is used to dissolve away the Crystalbond wax.
The individual devices are then mounted on a piece of pre-patterned
semi-insulating GaAs substrate. This substrate has ~20$\mu$m deep channels etched
and metallised to match the contact pads on the backgate side of the
device. Once the thinned device is positioned on the substrate,
small drops of silver epoxy are placed on these channels. The epoxy
flows along the channels and contacts the underside of the chip in
specific places. Once it is cured it provides electrical contact as
well as mechanical support to the thinned device. Gold-wire bonds
can then be made to the contact pads  on the topside
and to the the extensions of the backgates provided by the
conducting epoxy.

\section{A self-consistent model of gate action}
Techniques for calculating the band-bending and carrier densities in
layered heterostructures are well known. We review the procedure
very briefly. Normally a self-consistent solution of Poisson's
equation and the Schr\"odinger equation (for the envelop function)
is calculated\cite{stern-howard}. Poisson's equation relates the
electrostatic potential ($\phi(x)$) to the total charge density
$\rho(x)$
\begin{equation} \nabla \epsilon(x)\nabla\phi(x) =
-\rho(x)
\end{equation}

The envelope function ($\psi$) is determined by the Schr\"odinger
equation :
\begin{equation}
-\frac{\hbar^2}{2}\frac{d}{dx}(\frac{1}{m^{*}}\frac{d}{dx})\psi_j(x)
+ V(x)\psi_j(x) = E\psi_j(x)
\label{schrodingereq}
\end{equation}

Where $V(x) = |e|\phi(x) + {\Delta}E_c(x)$, the potential confining
the electrons is the sum of the electrostatic potential and the
conduction band offset at the heterointerface. $\epsilon(x)$ is the position dependent
dielectric constant, $E_j$ and $\psi_j$
are the subband energies and envelope functions that solve equation \ref{schrodingereq}. The
charge density at each point is given by the sum of the localised
charge ($N^{+}_{D}(x)$ on the ionised donors (or acceptors) and the
electron (hole) densities ($n(x)$) in the conduction (valence) band.
$E_D$ denotes the donor energy level.

\begin{eqnarray}
\rho(x)= && |e|N^{+}_{D}(x) - |e|n(x) \\
N^{+}_{D}(x)= && \frac{N{_D}(x)}{1 +
\frac{1}{2}\exp(\frac{E_f-E_D}{kT})}\\
n(x)=&&\sum\limits_{j}|\psi_j(x)|^2\frac{m^*}{\pi\hbar^2}\int_{E_j}^{\infty}\frac{dE}{1+\exp\frac{E-E_f}{kT}}
\end{eqnarray}

As long as no bias voltages are applied externally, the Fermi-level,
$E_f$, must be constant throughout the device and is generally taken
as the reference level. For a complete solution, the differential
equations require suitable boundary conditions. This may be provided
by the fact that $E_f$ is generally pinned near the middle
of the bandgap at the GaAs surface.\\
However, to understand the behaviour of a structure with external
voltages applied between two points ({\it e.g.} between an ohmic and
a gate or between two quantum wells), a further condition is necessary. The
external voltage sets the Fermi-level difference between the two
points. A calculation of $\rho(x)$, however requires that $E_f$ be
known at every point. A third condition, in addition to the two equations 1 \& 2,
is necessary, because we now need to solve for three variables
($\rho(x), \psi(x), E_f(x)$) at every point.\\
Thermodynamically, all net particle flows in a system can be traced
to a variation in the electrochemical potential \cite{marshak}. The Boltzmann transport equation yields
an expression relating  the net particle current $(\bf j)$ to
$E_f$ as

\begin{equation}
{\bf j}=n(x){\underline{\underline{\bf\mu}}} \nabla E_f
\end{equation}

In the relaxation time approximation ${\underline{\underline{\bf\mu}}}$ is given by
\begin{equation}
{\underline{\underline{\bf\mu}}}= \frac{e}{4\pi^3n}\int{\bf d^3k}\tau_e{\bf v_k}{\bf
v_k}\frac{\partial f^0}{\partial E} \label{mobilityeq}
\end{equation}
The derivative peaks sharply giving maximum weight to particles
at the Fermi surface. It is easy to show that if  $E({\bf k})$
is spherically symmetric, then the expression reduces to the simple
relation $\mu=e\tau/m^*$ at $T=0$.

The set of equations (1-6) are used to model current flow through
FETs under biasing\cite{kidong}, here we show that they can also be
used to \textit{approximately} model the band structure under a
voltage bias between the gate and the ohmic contacts with
intervening doped layers. In this case the current would correspond
(physically) to the very small leakage current that flows between
the gate and the conducting channels. We are not concerned with the
exact magnitude of this current. However, this current must be
constant, in a 1-dimensional case. Thus the variation of $E_f$ would
be slow in regions where the product of $n\mu$ is large and most of
the ``potential drop" would take place in regions where $n\mu$ is
smaller as  would be expected in analogy with
voltage drops across unequal resistances in series. The assignment
of a mobility to the carriers at each point can only be done
approximately, using a mobility model, rather than an evaluation of
the relaxation time average, following equation \ref{mobilityeq}. A
``mobility-model" is an empirical relationship between the bulk-doping,
electric field strength, temperature and mobility. First proposed in
the context of Silicon\cite{caughey}, it has been adopted for use
with various other semiconductors. The use of an approximate bulk
mobility may seem unusual at first glance, however several other
bulk parameters- like dielectric constant, effective mass and
donor/acceptor ionisation energies are used in calculating the
band-bending in heterostructure based devices.  Also, in regions
where the error from the mobility term may be the highest, it gets
multiplied by a very small carrier density, thus reducing the net
effect of the error on the calculation. It is also important to note
that the bulk semiconductor could be in a metallic or insulating
regime depending on the doping concentrations. No effort is made
here to take into account the mechanisms behind the conduction
process, but an empirical relation is utilised. Finally we show that
the use of this approximate procedure leads to correct predictions
and describes experimentally observed temperature dependent
behaviour of gated
structures correctly.\\
The specific mobility model\cite{caughey,kidong} used here (SIMBA) assumes
that
\begin{equation}
\mu_{n,E0}(N_D)=\mu_{n,min} + \mu_{n,D}/[1 + (N_D/N_{n,ref})^{\alpha_n}]
\end{equation}
\begin{equation}
\mu_{n}(\vec{E})=\mu_{n,E0}/
[1 + (\mu_{n,E0}{\large \frac{|\vec{E}|}{\nu_{n,S}}})^{\kappa_n}] ^{1/\kappa_n}
\end{equation}
\begin{equation}
\nu_{n,S}=\nu_{n,0} - d_{n,V}(T-T_0)
\end{equation}
\begin{equation}
\mu_n(T)=\mu_n(T/T_0)^{-\gamma^n}
\end{equation}
The empirical parameters
$\mu_{n,min}$=$1000$ ($32$)  cm$^2$V$^{-1}$s$^{-1}$ ,
$\mu_{n,D}$ = $7200$ ($400$) cm$^2$V$^{-1}$s$^{-1}$,
$\alpha_n$=0.55 (0.5),
$N_{n,ref}$ = $6\times10^{16}$ ($1.88\times10^{17}$) cm$^{-3}$
$\kappa_n$ = 4 (4),
$\gamma^n$ = 1 (2.1),
$\nu_{n,0}$ = $1.5\times10^7$ ($0.5\times10^7$) cms$^{-1}$,
$d_{n,V}$ = $1.5\times10^4$ ($1.5\times10^4$) cms$^{-1}$K$^{-1}$
and $T_0$ = 300 (300) K. The values within the braces denote the corresponding numbers for
p-type material.

We have used an academic version of the software
``nextnano$^3$"\cite{nextnano} (developed at the Walter Schottky
Institute, Munich) for solving the Poisson-Schr\"{o}dinger-current equation self
consistently.

\subsection{Temperature dependence of gate action on a simple HEMT}
\begin{figure}[h]
\includegraphics[width=7.75cm,clip]{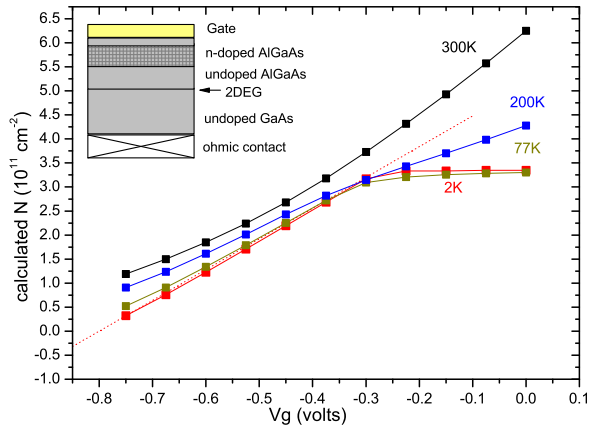}
\caption{\label{surfacegateaction} (Color online) Figure shows
(a) modeling of the action of a surface gate at different temperatures.
The calculated traces match data from Hirakawa {\it et al}\cite{hirakawa},
very closely}
\end{figure}

Using the equations stated in the previous section
we have calculated the carrier density at a
heterointerface as a function of the  surface gate voltage at
various temperatures (see fig \ref{surfacegateaction}a). Usually the
behaviour of a device with ohmics (source and drain contacts) and a
gate needs to be treated as a two-dimensional structure. We find
that a simplification can be used for our purposes that allows us to
model this as a one-dimensional problem.
Fig \ref{surfacegateaction}a shows the placement of an ohmic contact
at one end of the device and a surface gate on the other. The
voltage bias is then applied between these two. The simplification
has its limitations, but it still allows us to deduce some important
conclusions while designing wafers. Experimental data from Hirakawa et al,
\cite{hirakawa} on the temperature dependence of the gate action
on a  HEMT structure shows certain features which a simple
capacitor model of a surface gate would not explain.
The calculation correctly predicts  the observed features
(Fig 2 of ref\cite{hirakawa}). In particular
we are able to explain the slight upward curvature of the
carrier-density vs gate-voltage trace at low gate biases and high
temperatures. The slight flattening of the traces at low
temperatures is correctly reproduced. As far as we know, this
feature of a simple gated device has not been explained clearly
before. According to these calculations there is some transfer of
charge to the impurity band as long as the impurity band ($E_D$
below the conduction band in the doped AlGaAs) is in contact with
the electrochemical potential. As the negative bias on the gate is
increased, the conduction band and the impurity band are both pulled
up and at one point moves fully above the local electrochemical
potential. If the doping is higher this point would shift to more
negative gate voltages. It is at this point that the gate starts
acting on the 2DEG and the variation of the carrier density is fully
accounted for by the gate to 2DEG capacitance. The transfer of
charge into dopant states requires that there be some residual
hopping-conduction in the impurity band \cite{mott}. This condition
is not very difficult to satisfy (without considering tunnelling
from surface states etc.), because the metal-insulator transition
point in bulk-GaAs lies in the doping density range of $\sim
10^{16} - 10^{17}$cm$^{-3}$.\\

\subsection{Maximum doping density in inverted structures for
backgate action} 
\begin{figure}[h]
\begin{center}
\includegraphics[width=9cm,clip]{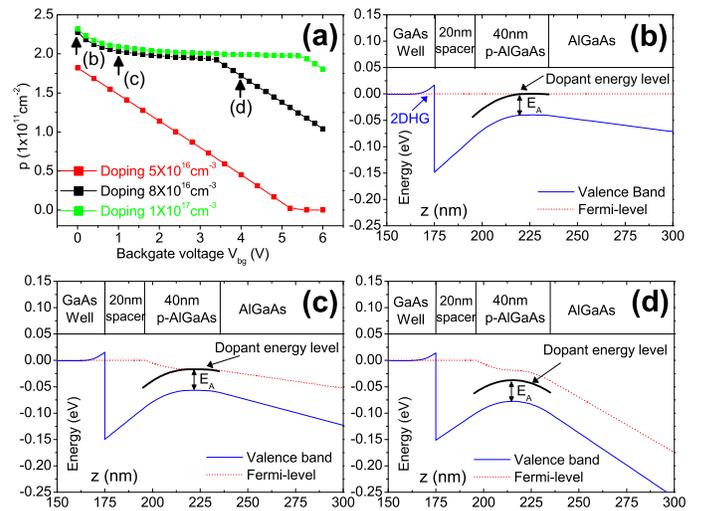}
\end{center}
\caption{\label{invHHMTbias} (Color online) Positions of the dopant
energy levels and the electrochemical potential under biasing.
Figures (b),(c),(d) correspond to positions marked (b),(c),(d) with arrows in
fig (a) for the $8\times 10^{16}$cm$^{-3}$ doping. }
\end{figure}

This procedure also allows us to estimate the
maximum doping density at which a surface gate (or a backgate) would
work. We use this to design an inverted hole gas that may be gated
from below in a thinned sample as described in section II. An
interesting observation that emerges, is that for inverted
structures the doping has to be significantly less (compared to the
structures described in fig \ref{surfacegateaction}  for a gate to
work. The reason for this is that the distance of the dopants from
the substrate interface in most inverted structures is much greater
than the distance of the dopants from the surface in the ``normal"
structures. This causes the curvature of the bands, around the doped
regions, in inverted structures to be much less and consequently,
due to this ``flatness" the dopant band cannot move away from the
local electrochemical potential unless the doping itself (hence the
density of dopant states) is considerably reduced.\\
Fig \ref{invHHMTbias} shows the (simulated) effect of a bias on
devices  doping densities $1\times10^{17}$cm$^{-3}$,
$8\times10^{16}$cm$^{-3}$, $5\times10^{16}$cm$^{-3}$. The gate begins to
work around a doping density of $8\times10^{16}$cm$^{-3}$ and when the
hole density is just below $2\times10^{11}$cm$^{-2}$. The position of
the impurity band with respect to the local electrochemical
potential is shown in fig\ref{invHHMTbias}. Comparison with data from measured
devices show that the action of the backgate is linear in devices
with a starting density of $2\times10^{11}$cm$^{-2}$ or lower, as
inferred from the calculations. However the doping density at which
this happens appears to be higher (in measured devices) by nearly a
factor of 2.

\section{Fabrication of an independently contacted tunable
electron-hole bilayer} Interest in designing an electron-hole
bilayer stems from the fact that the attractive Coulomb interaction
between electrons and holes spaced $\sim$20nm apart in GaAs double
quantum well (DQW) structures can give rise to novel excitonic
phases\cite{lozovik, littlewood, vignale}. Excitons are bosonic
entities with a very low effective mass. A collection of excitons
with high enough density may be expected to undergo a Bose-Einstein
condensation at a much higher temperature (as high as $\sim$ 1K),
than a  heavy inert gas or alkali atom cloud, where condensation has been
observed at $\sim$ 200$\mu$K at the highest. However, it is now well
understood that in DQW structures using the GaAs-AlGaAs system the
electrochemical potential of the electrons and holes must differ by
approximately 1.5V - the bandgap of GaAs. If a very closely spaced 2DHG and 2DEG are to be supported
without any interlayer bias, a constant electrochemical potential  would
have to cut the conduction band and the valence band within 15-20nm. The necessary band bending would
imply an electric field of $10^8$V/m, which is far more than the breakdown field
of GaAs. However to bias a layer with
respect to another layer, one needs independent ohmic contacts to
both layers. Additionally the narrow interlayer barrier must be able
to withstand the bias with very little leakage.  Though excitonic phases in
these structures have been predicted many years back\cite{lozovik}, but
experimentally such devices have proved extremely difficult to make.
Even with the present day developments of MBE techniques, these
devices appear to be on the borderline of feasibility. However the
possibility of excitons with infinite lifetime, leading to phenomena
like dipolar superfluidity, BEC etc have led to quite a few
experimental attempts in recent
years\cite{sivan,pohlt,keogh,seamons}. Here we apply the backgating
technique presented in section I, to design a bilayer device in
which the two layers are independently contacted. The two densities
can be tuned independently, within limits, by using a backgate and
an interlayer voltage. Here we present data from three devices.

\begin{figure}[h]
\begin{center}
\includegraphics[width=5cm,clip]{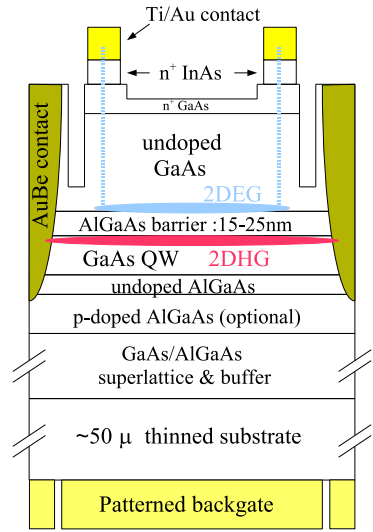}
\end{center}
\caption{\label{sampleschematic} (Color online) Generalised schematic of the electron-hole devices. Certain details like the
doping levels, width of the quantum wells, width and composition of the barriers very between devices. See table 1 for these
details }
\end{figure}

\begin{figure}[h]
\begin{center}
\includegraphics[width=9cm,clip]{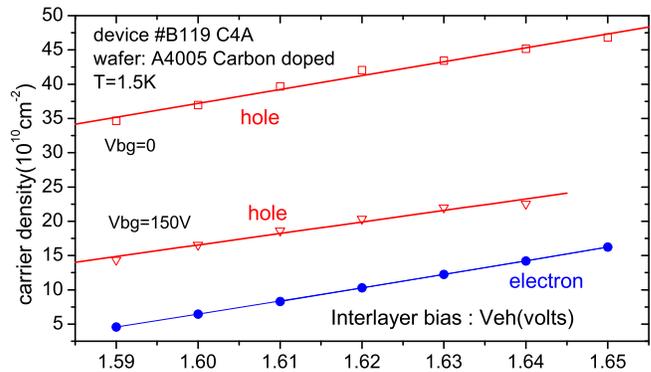}
\end{center}
\caption{\label{backgateaction} (Color online) Action of backgate. Data from device A. The electron and hole
densities were not matched in this device. The electron densities are shown for $V_{bg}$=0. The change in the
electron density with backgate voltage is very small.}
\end{figure}

In device A and B (see fig \ref{sampleschematic} for a schematic) an inverted hole gas was created by
modulation doping (Carbon in AlGaAs) using the estimates
presented in the previous section. Device C was completely undoped and the
hole gas was created by biasing the backgate to attract holes below the contacts.
Considering the dielectric constant of GaAs, $\epsilon/\epsilon_0\approx12.8$, a gate
50-60$\mu m$ away from the quantum well is expected to be able to change
(either deplete or induce) carrier densities by $\approx6\times10^{10}$cm$^{-2}$ for every 50 volts
of biasing. The measured changes in the carrier densities were close to this expected number.\\
For example in the data from device A (see fig \ref{backgateaction}), the hole density changes by
$\Delta p\approx2\times10^{11}$cm$^{-2}$ for a backgate bias of 150V.
AuBe contacts alloyed at 500$^o$C were used to contact the 2DHG.
In device C (fully undoped), the backgate has two discontinous parts. The part below the p-type contacts
is biased negative to attract holes, the part below the central region (which overlaps the 2DEG), is usually
biased positively to deplete excess holes. There is a 20$\mu$m gap between the two gates (see fig \ref{process-steps-fig}).
Contact to the overlapping
2DEG was made by using a different technique\cite{keogh}. The top layer of the wafer consisted of
n$^+$ InAs ($8\times10^{18}$cm$^{-3}$ Si doping). The surface states of InAs have the unusual property of pinning the Fermi level {\em above}
the bottom of the conduction band and not at the midgap like the surface states of GaAs. Thus any metal
deposited on a clean InAs surface would not see a barrier for electron injection. It can be interpreted as
a negative Schottky barrier and does not require any post deposition annealing. Using a selective etch (dry conc. HCl) the InAs is removed from all places except where contacts to the electron layer are required. Since this contacting
process requires no annealing, the contact material does not spike into the underlying semiconductor and is
able to form a low resistance shallow contact to the 2DEG, without shorting to the 2DHG.

\begin{table}[h]
\caption{Summary of the three devices. All the three devices have an approximately 1$\mu$m thick GaAs/AlGaAs superlattice near
the substrate.}
\begin{tabular}{|l|c|c|c|c|}
\hline
device & p-doping                                    & \parbox{1cm}{\vspace*{10pt}QW \\width\vspace*{10pt}} & barrier & \parbox{1.5cm}{ matched densities possible} \\ \hline
A      & \parbox{2.2cm}{\vspace*{10pt} $1.5\times10^{17}$cm$^{-3}$ \\ $\times$ 40nm \\wafer ID\\A4005\vspace*{10pt}} & 40nm & \parbox{2cm}{25nm \\ Al$_{0.3}$Ga$_{0.7}$As} & No  \\ \hline
B      & \parbox{2.2cm}{\vspace*{10pt}$5\times10^{16}$cm$^{-3}$  \\ $\times$ 40nm\\wafer ID\\A4170\vspace*{10pt}}   & 20nm & \parbox{2cm}{15nm \\ Al$_{0.9}$Ga$_{0.1}$As}  & Yes \\ \hline
C      & \parbox{2.2cm}{no doping\\wafer ID\\A4142}    & 20nm & \parbox{2cm}{\vspace*{10pt}25nm \\ Al$_{0.3}$Ga$_{0.7}$As\vspace*{10pt}}  & Yes \\ \hline
\end{tabular}
\end{table}

\begin{figure}[h]
\begin{center}
\includegraphics[width=9cm,clip]{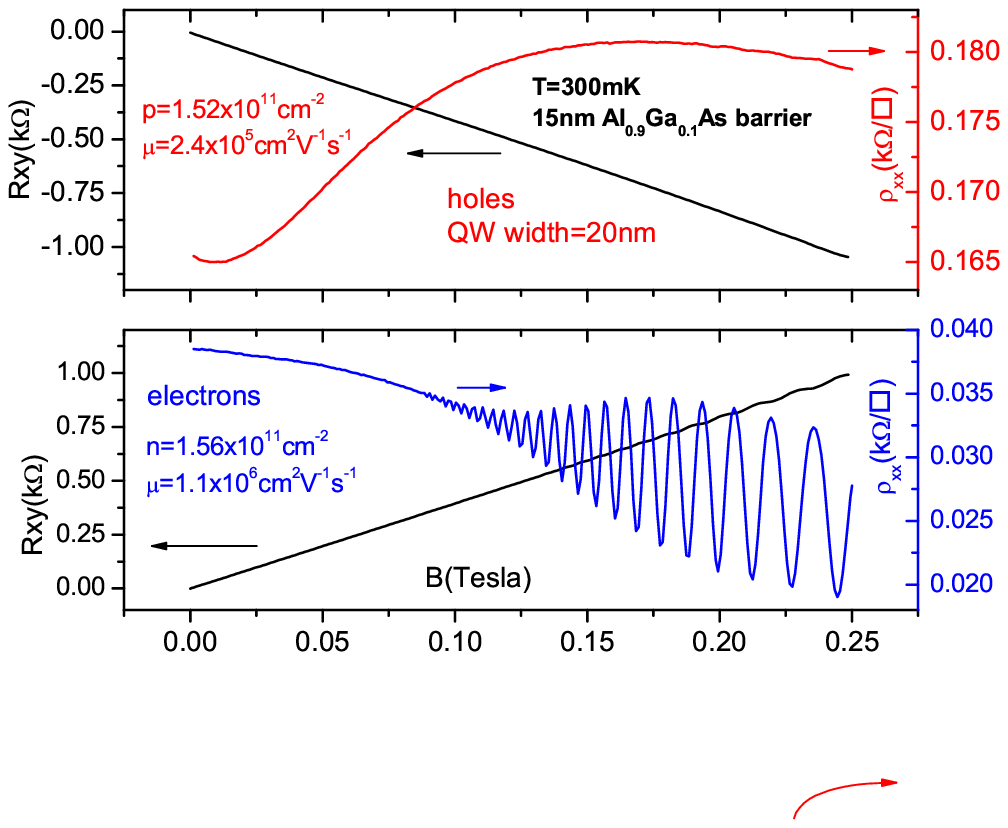}
\end{center}
\caption{\label{npmatched1_5e11} (Color online) Data from device B, n=p=$1.5\times10^{11}$cm$^{-2}$}
\end{figure}

\begin{figure}[h]
\begin{center}
\includegraphics[width=9cm,clip]{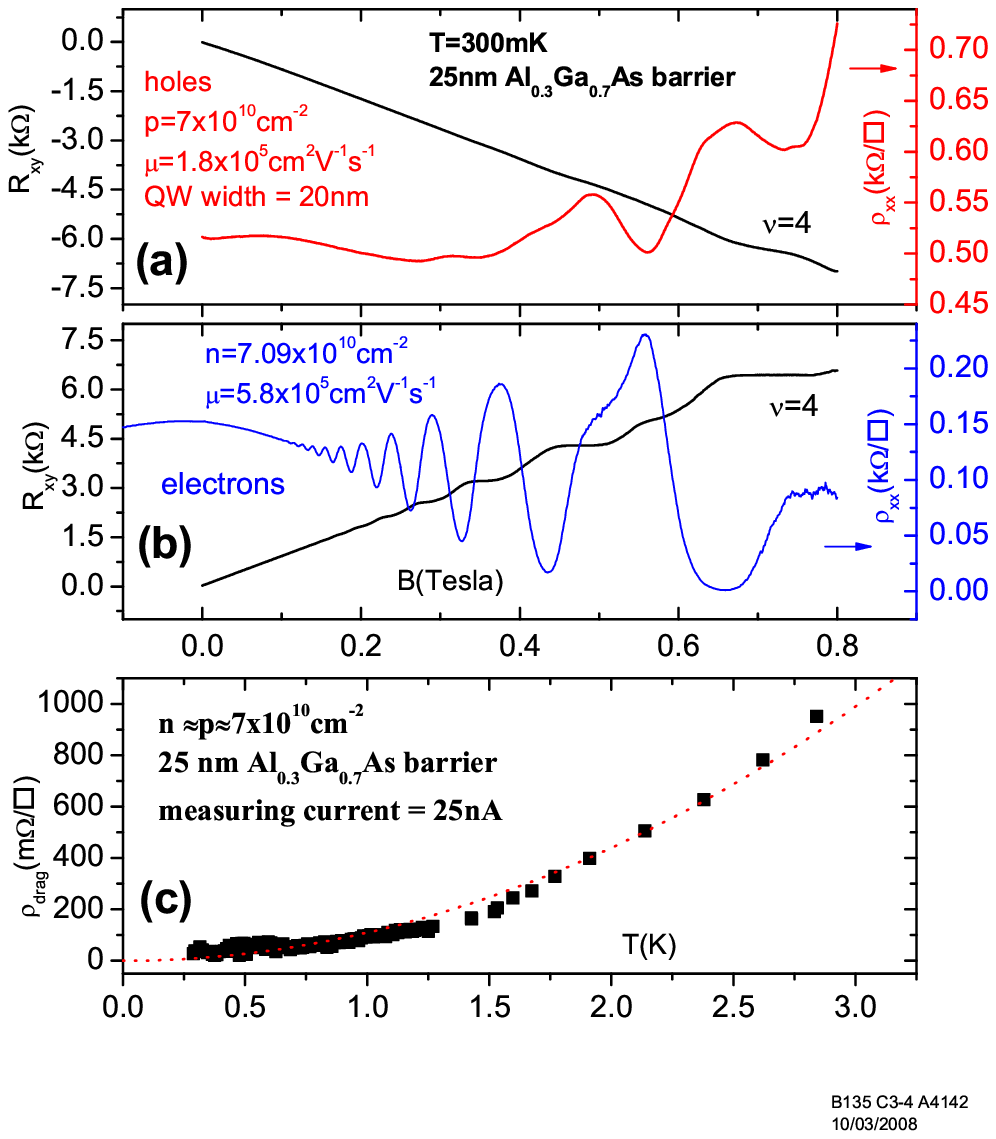}
\end{center}
\caption{\label{npmatched7e10} (Color online) (a) \& (b): Data from device C, n=p=$7\times10^{10}$cm$^{-2}$.
The lowest panel (c) shows drag resistivity measured by passing current through the hole layer and
measuring the voltage across the electron layer.}
\end{figure}

A typical variation of the electron and hole densities with the interlayer bias ($V_{eh}$) and the backgate bias ($V_{bg}$) is
shown in fig \ref{backgateaction}. The comparatively higher doping in this device led to a higher hole density.

Comparing the electron and hole mobilities in the undoped device (C) at $n=p=7\times10^{10}$cm$^{-2}$ we find that $\mu_e/\mu_h$=3.2.
Assuming an effective mass ratio $m^*_h/m^*_e\approx7.5$, (appropriate for heavy holes at low densities)
we find that the transport scattering times $\tau_h/\tau_e\approx2$.
This probably results from the doped surface  contributing to scattering in the electron layer more than it does to the hole layer.
In addition hole wavefunctions tend to be smaller than electrons, leading to an overlap with
 a smaller number of background impurity sites. Both the 2DEG and the 2DHG otherwise sees a similar background impurity density.
 A comparison of the mobilities in device C at $n=p=1.5\times10^{11}$cm$^{-2}$  gives $\mu_e/\mu_h$=4.6.
 Calculations of $E_{hole}$ vs $k_{||}$ for the holes using $\vec{k}.\vec{p}$ method show no significant
 non-parabolicity or mixing of light and heavy holes at densities less than $p=2\times10^{11}$cm$^{-2}$ for a 20nm QW\cite{winkler}.
 These calculations are not presented here but we continue to assume a parabolic heavy hole band for these estimates and find that
  $\tau_h >\tau_e$, though the difference appears to decrease at higher densities\cite{cpmorath}.

In these devices, we are also able to directly measure the
interlayer interaction by the Coulomb drag method\cite{pogrebinsky}.
These measurements have been performed on some of the devices at
temperatures down to 300mK but have not yet shown  unambigous
signatures of an excitonic phase. However regimes of stronger
electron-hole interaction can be achieved in these devices, by
reducing the densities and the barrier width further.
It is expected that these  devices would prove useful in  transport
based studies of bilayer excitonic phases.

\section{Conclusion}
In conclusion, we have developed a method to achieve the
functionality of a double-sided mask aligner using a much less
expensive and more commonly available single-sided aligner. We have used
this technique in conjunction with a numerical procedure to estimate
the behaviour of the electrochemical potential in doped and
(back)gated devices under biasing, to design an electron-hole
bilayer with independent contacts and a patterned backgate. In these systems, the
interlayer scattering rate between electrons and holes is of considerable interest.
The scattering rate may show some very interesting characteristics at low temperatures and
densities.  Our alignment method and calculation procedure are
however applicable to a much wider variety of semiconductor devices.

\section{Acknowledgements}
This work was funded by EPSRC, UK. IF would like to thank Toshiba Research Europe Ltd. for financial support.
MT acknowledges support from the Gates Cambridge Trust.

\end{document}